\documentclass[%
 reprint,
 amsmath,amssymb,
 aps,
]{revtex4-1}
\usepackage{url}
\usepackage{graphicx}
\usepackage{dcolumn}
\usepackage{bm}

\begin{document}

\title{Effectively Trainable Semi-Quantum Restricted Boltzmann Machine}

\author{Ya.S. Lyakhova$^{1,2,3}$}
\email{yanalyakhova@gmail.com}
\author{E.A. Polyakov$^1$}
\author{A.N. Rubtsov$^{1,2}$}
\affiliation{$^1$Russian Quantum Center, Skolkovo Innovation city, 121205 Moscow, Russia \\
$^2$NTI Center for Quantum Communications, National University of Science and Technology MISiS, 119049 Moscow, Russia\\
$^3$National Research Nuclear University MEPhI, 115409 Moscow, Russia
}

\date{\today}

\begin{abstract}
We propose a novel quantum model for the restricted Boltzmann machine (RBM), in which the visible units remain classical whereas the hidden units are quantized as noninteracting fermions. The free motion of the fermions is parametrically coupled to the classical signal of the visible units. This model possesses a quantum behaviour such as coherences between the hidden units. Numerical experiments show that this fact makes it more powerful than the classical RBM with the same number of hidden units. At the same time, a significant advantage of the proposed model over the other approaches to the Quantum Boltzmann Machine (QBM) is that it is exactly solvable and efficiently trainable on a classical computer: there is a closed expression for the log-likelihood gradient with respect to its parameters. This fact makes it interesting not only as a model of a hypothetical quantum simulator, but also as a quantum-inspired classical machine-learning
algorithm.
\end{abstract}

\maketitle

\section{Introduction}

Nowadays machine learning becomes an all-pervasive paradigm of how to obtain, process and store knowledge. Initially arising in the field of computer science, it finds numerous interdisciplinary applications. They range from commercial applications like speech and handwriting recognition,  classification/recognition of video content \cite{Taylor2006}, up to various scientific applications. As an example of the latter, in physics the most prominent applications are the generative modeling of quantum control and measurement protocols \cite{Niu2019}, of condensed matter problems \cite{Carrasquilla2017}, and of quantum systems \cite{Carleo2017}.

One of the major approaches to machine learning is the generative modeling. Here, given a certain finite data set $X$, a model of its probability distribution $P\left(X\right)$ is estimated. The tuning of the model parameters (i.e. training)  is usually carried over by maximizing a certain distribution-resemblance measure. There are two criteria of a successful learning model. Firstly, it has to be easily trainable. It turns out that too complex models (or machines) are challenging and impractical to train. Stacks of simple models is a way round this problem. While each layer is easy to train, their composition can model data of high statistical complexity \cite{LeCun2015}. The second criterion is imposed by the problem of overfitting. If one increases the number of model parameters for a given data set, the model tends to approximate the particular realization of the random data scatter thus losing its predictive capability. It entails that a successful machine has to be simple enough to avoid overfitting.

In the classical machine learning, Restricted Boltzmann Machine (RBM) represents such a model. RBM is a two-layer energy model with no coupling between the units of the same layer. One layer is called visible and represents the observable data, while the other one is called hidden and represents some statistical correlations between the visible units. The energy of such a machine is a quadratic function of the visible and hidden variables, and their probability distribution is given by the finite-temperature Boltzmann distribution \cite{Hinton2010}. The absence of connections between the units in the same layer makes RBM a highly efficiently trainable model. Originally, the parameters of RBM, namely as biases and weights, are assumed to be real. Nevertheless, recently a complex-valued RBM was proposed \cite{Carleo2017}. Such an extension allows the machine to treat quantum properties of the system it models, namely the wave-function amplitude and its phase.

We live in the second quantum revolution era, which is characterized by experimental and technological achievements in individual quantum systems control. The major motivation behind this activity is to devise a feasible quantum computing circuits, which could outperform the capabilities of classical computing devices \cite{Nielsen2002}. The field of machine learning does not stand aside. Quantum models of machine learning are being proposed \cite{Rebentrost2014, Rebentrost2018}, and in particular, recently the quantum Boltzmann machine (QBM) and its restricted version (RQBM) were proposed \cite{Amin2018}. The main difference from the classical Boltzmann machine is that both the visible and hidden units (spins) are allowed to be in a quantum superposition. The energy function (Hamiltonian) is modified to include a non-diagonal connections of spins to an external field (bias). While RQBM was demonstrated to learn the data distribution better than the classical RBM, its major drawback is that this model is of high computational complexity, and its training is rather challenging. 

The purpose of this work is to present a quantum version of the Restricted Boltzmann Machine, which at the same time has reasonable computational complexity so that it could be efficiently trained. We propose to represent the hidden units by non-interacting fermions. Technically it amounts to the replacement of a vector of $m$ hidden units with a square matrix of the size $m\times m$, and thus hidden bias and coupling between the layers (weights) are now represented by $m\times m$ complex-valued matrix and $n\times m\times m$ complex-valued matrix respectively, for $n$ visible units. Analogously to the classical RBM, where the coupling between the layers is linear, the free-motion Hamiltonian for the fermionic hidden units is linearly coupled to the visible (classical) units. We call it semi-quantum Restricted Boltzmann Machine (sqRBM).

We evaluate the proposed sqRBM model on two training data sets: the ensemble of Bars\&Stripes \cite{Fischer2010}, and the Optdigits data set \cite{OptDigits}.  We use log-likelihood as the distributions proximity measure to compare sqRBM with the classical RBM. We also present the results of the cross-validation test to compare the overfitting of the machines.
 
In the Section 2 we briefly overview the RBM model and its training procedures. Section 3 is devoted to the detailed description of the proposed semi-quantum RBM. In this case the probability distribution is given in terms of density matrix, and the usual summation over the classical hidden units is replaced by the taking of trace w.r.t. quantum fermionic hidden units. We also derive learning rules for the training via gradient ascent algorithm, and discuss the numerical implementation in details. In the Section 4 we show the results of the numerical experiments conducted with the use of Bars\&Stripes and Optdigits data sets. Our conclusions are set out in the Section 5.

\section{Classical Restricted Boltzmann Machine}
\subsection{Model}

\begin{figure}
    \begin{minipage}{0.47\textwidth}
        \includegraphics[scale=0.3]{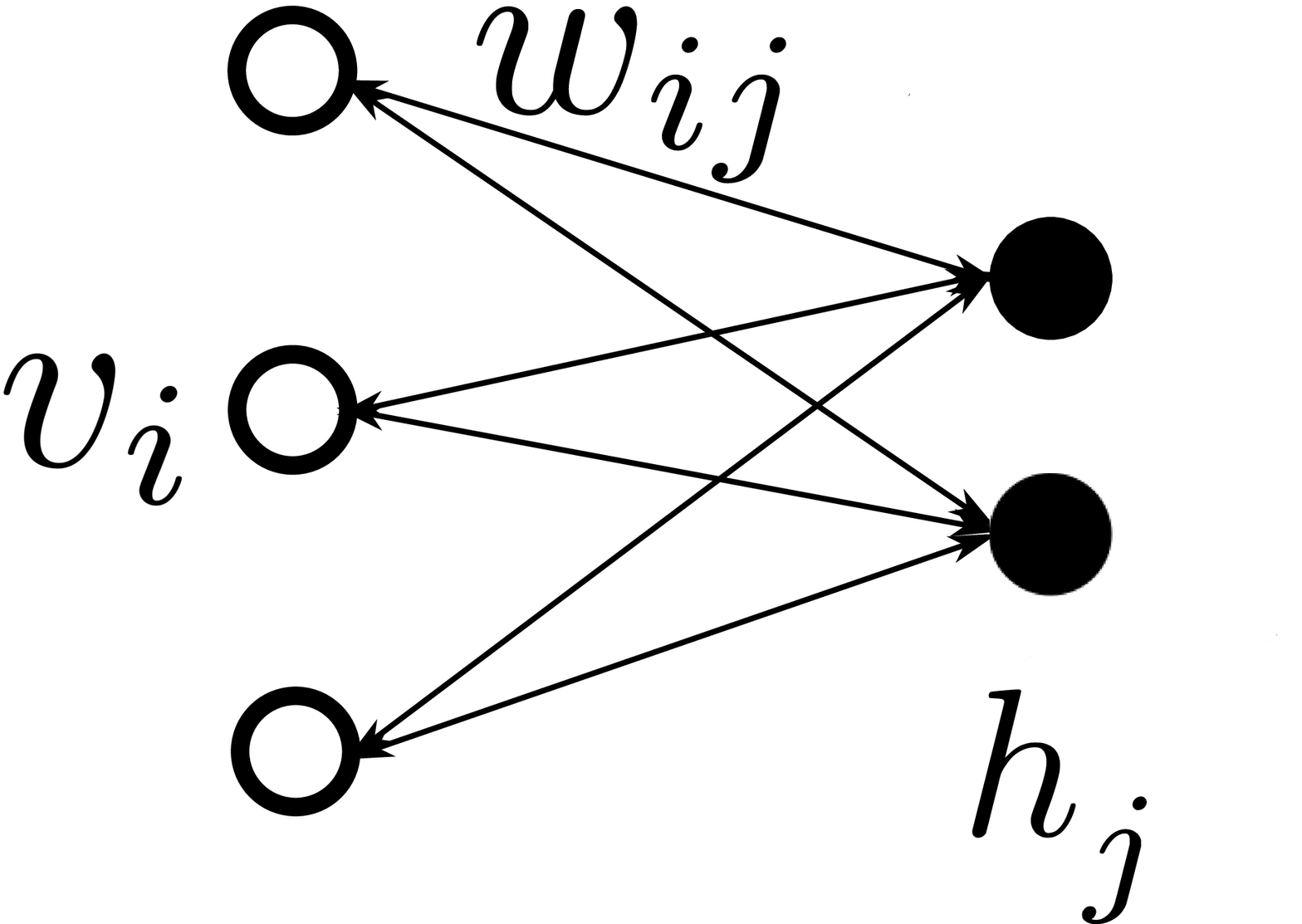} \\ (a)
    \end{minipage}
    \hfill
    \begin{minipage}{0.47\textwidth}
        \includegraphics[scale=0.3]{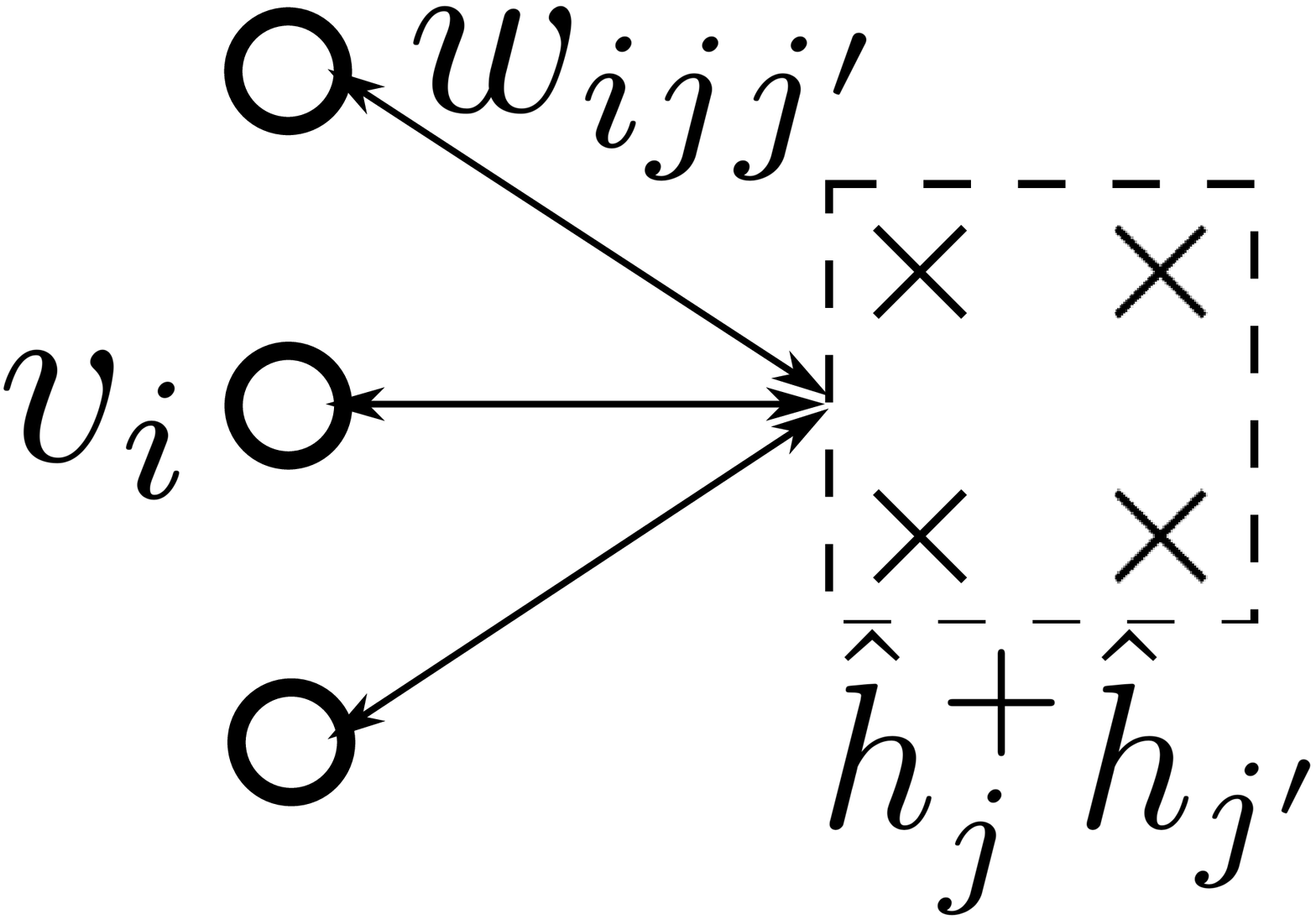} \\ (b)
    \end{minipage}
    \caption{RBM (a) and sqRBM (b) schemes. They consist of two interacting layers, one of which is called visible $v_i$ and the other is called hidden. Dimensionality of sqRBM hidden layer $\hat{h}_j\hat{h}_j^{\prime}$ is twice the dimensionality of the corresponding RBM hidden layer $h_j$. The interaction between the layers is denoted with $w_{ij}$ for the RBM and with $w_{ijj^{\prime}}$ for the sqRBM.}
    \label{fig:scheme}
    \end{figure}

Restricted Boltzmann Machine (RBM) is a bipartite undirected neural network (see figure \ref{fig:scheme}(a)). One of it's layers is conventionally called visible and the other one hidden (see figure \ref{fig:scheme}(b)). Both visible ($\mathbf{v}$) and hidden ($\mathbf{h}$) neurons are assumed to take the values $\{0, 1\}$. The former is used to describe the observable data, whereas the latter is used to model the correlations between the observable components.

RBM is an energy-based model, which dynamics can be described by the Gibbs-Boltzmann distribution:
\begin{equation}
    p(\mathbf{v}, \mathbf{h}) = \frac{1}{Z}e^{-E(\mathbf{v}, \mathbf{h})},
\end{equation}
with the conventional definition of partition function
\begin{equation}
    Z = \sum_{\mathbf{v}}\sum_{\mathbf{h}} e^{-E(\mathbf{v}, \mathbf{h})}.
\end{equation}
Physically it corresponds to the assumption that RBM system is in the thermal equilibrium at the finite temperature $1/T = 1$. The energy $E(\mathbf{v}, \mathbf{h})$ of the machine is postulated to be a quadratic function of it's variables. For the RBM of $n$ visible units and $m$ hidden ones we have
\begin{equation}
\label{eq:RBMen}
    E(\mathbf{v}, \mathbf{h}) = - \sum_{i = 1}^nb_iv_i - \sum_{j=1}^mc_jh_j - \sum_{i=1}^n\sum_{j=1}^mv_iw_{ij}h_j.
\end{equation}
Here, coefficients $\mathbf{b}$ and $\mathbf{c}$ are called biases, whereas $\mathbf{w}$ is a weight matrix.

The absence of connections between the units of the same layer of RBM allows one to treat their states as mutually independent in terms of probability, namely \cite{Fischer2012}
\begin{equation}
    p(\mathbf{h}|\mathbf{v}) = \prod_{j=1}^mp(h_j|\mathbf{v}),
\end{equation}
and vice versa. This property makes RBM solvable and effectively trainable model via the method of direct Gibbs sampling \cite{Hinton2002}.

\subsection{Training}
The main idea of RBM usage is to model the distribution of observable data $p^{data}(\mathbf{v})$. This is achieved by the adjustment of the biases $\mathbf{b}, \mathbf{c}$ and weights $\mathbf{w}$ (the parameters of the model) so that the marginal distribution of the visible layer $p(\mathbf{v})$ would be as close as possible to the target distribution $p^{data}(\mathbf{v})$.

The measure of target and model distributions proximity can be represented by the log-likelihood function
\begin{equation}
\label{eq:lhood}
    \mathcal{L}(\mathbf{b},\mathbf{c},\mathbf{w}) = \frac{1}{N_{data}}\sum_{\mathbf{v}_{data}} \log P(\mathbf{v}_{data}).
\end{equation}
The higher $\mathcal{L}(\mathbf{b},\mathbf{c},\mathbf{w})$ the better given RBM models the observables' distribution.

Maximizing the log-likelihood $\mathcal{L}$ (or equivalently minimizing the negative log-likelihood $-\mathcal{L}$) can be implemented by the gradient ascent (descent) algorithm \cite{Hinton2010}. It can be shown \cite{Fischer2012} that it gives the following update rules for biases and weights:
\begin{subequations}
\label{eq:rbm_grads}
\begin{eqnarray}
    \Delta \mathbf{b} &=& \eta \partial_{\mathbf{b}}\mathcal{L}(\mathbf{b},\mathbf{c},\mathbf{w}) = \nonumber \\
    &=&\eta \left( \frac{1}{N_{data}}\sum_{\mathbf{v}_{data}}\mathbf{v}_{data} - \sum_{\mathbf{v}}P(\mathbf{v})\mathbf{v} \right),
\end{eqnarray}
\begin{eqnarray}
    \Delta \mathbf{c} &=& \eta \partial_{\mathbf{c}}\mathcal{L}(\mathbf{b},\mathbf{c},\mathbf{w}) = \nonumber \\&=& \eta \left(  \frac{1}{N_{data}}\sum_{\mathbf{v}_{data},\mathbf{h}}P(\mathbf{h}|\mathbf{v}_{data})\mathbf{h} \right.\nonumber \\
    &&\qquad\qquad\left.- \sum_{\mathbf{v},\mathbf{h}} P(\mathbf{h}|\mathbf{v})\mathbf{h} \right),
\end{eqnarray}
\begin{eqnarray}
    \Delta \mathbf{w} &=& \partial_{\mathbf{w}} \mathcal{L}(\mathbf{b},\mathbf{c},\mathbf{w}) = \nonumber \\&=& \eta \left( \frac{1}{N_{data}} \sum_{\mathbf{v}_{data}, \mathbf{h}}P(\mathbf{h}|\mathbf{v}_{data})\mathbf{v}_{data}\mathbf{h} -\right. \nonumber \\
    &&\qquad\qquad\left.-\sum_{\mathbf{v},\mathbf{h}}P(\mathbf{h}|\mathbf{v})\mathbf{v}\mathbf{h} \right).
\end{eqnarray}
\end{subequations}
Here $\eta$ is called learning rate and defines the size of a single gradient ascent step.

There are various techniques of how to actually evaluate the gradients (\ref{eq:rbm_grads}). In this work we employ the fast mean-field-like algorithm called Contrastive Divergence (CD) \cite{Hinton2002}, and the full Monte-Carlo simulation called Persistent Contrastive Divergence (PCD) \cite{Tieleman2008}.
\section{Semi-Quantum Restricted Boltzmann Machine}
\label{sec:SFM}

\subsection{Model}

We propose here semi-quantum Restricted Boltzmann Machine (sqRBM), which represents a two-layer system (see figure \ref{fig:scheme}(b)). Its visible layer consists of classical binary variables $\mathbf{v}$, whereas the hidden one is composed of quantum degrees of freedom $\hat{\mathbf{h}}$. In the spirit of Schwinger-Wigner representation each hidden unit is represented by a pair of fermionic creation/annihilation operators. Such a system is described by the following Hamiltonian:

\begin{eqnarray}
\label{eq:ham}
    \hat{H}(\mathbf{v}, \mathbf{h}) = - \sum_{i = 1}^n b_iv_i - \sum_{j,j^{\prime}=1}^mc_{jj^{\prime}}\hat{h}_j^+\hat{h}_{j^{\prime}} \nonumber \\
    - \sum_{i=1}^n\sum_{j,j^{\prime}=1}^mv_iw_{ijj^{\prime}}\hat{h}^+_{j}\hat{h}_{j^{\prime}},
\end{eqnarray}
where $\mathbf{c}$ is a Hermitian matrix and $w_{ijj^{\prime}}$ is a Hermitian matrix w.r.t. $j,j^{\prime}$ for a given $i$.

As in the classical case we assume that sqRBM is in the state of thermal equilibrium at some finite inverse temperature $\beta$. Thus its state is described by the canonical ensemble in terms of the density matrix
\begin{equation}
    \hat{\rho}(\mathbf{v}, \mathbf{h}) = \frac{e^{-\beta \hat{H}(\mathbf{v}, \mathbf{h})}}{Z}.
\end{equation}
Here the partition function $Z$ of this hybrid classical-quantum system is given by\begin{equation}
    Z = \sum_{\mathbf{v}}\mathrm{Tr}_h(e^{-\beta\hat{H}(\mathbf{v,\mathbf{h}})}),
\label{eq:Z}
\end{equation}
where $\sum_{\mathbf{v}}$ stands for the ordinary summation over the observable classical units $\mathbf{v}$, and the trace $\mathrm{Tr}_h$ is taken over the quantum hidden degrees of freedom. Hereinafter we set $\beta = 1$ for simplicity.

Fermionic hidden subsystem obeys the Fermi-Dirac statistics:
\begin{equation}
\label{eq:FermiDirac}
    p(\hat{h}^+_j\hat{h}_{j^{\prime}}|\mathbf{v}) = \left( \frac{1}{1 + e^{-H_h}} \right)_{jj^{\prime}}, 
\end{equation}where $H_h = - c_{jj^{\prime}} - \sum_{i=1}^nv_iw_{ijj^{\prime}}$ is a $m \times m$ matrix. It is clear now that in the case when the Hamiltonian (\ref{eq:ham}) is diagonal w.r.t. the fermionic degrees of freedom, namely $c_{jj^{\prime}} = c_{jj}\delta_{jj^{\prime}}$ and $w_{ijj^{\prime}} = w_{ijj}\delta_{jj^{\prime}}$, it is fully equivalent to the classical RBM (\ref{eq:RBMen}).

\subsection{Training}
As in the case of classical RBM, the goal of training is to approximate the probability distribution of the input data set $\tilde{p}(\mathbf{v})$ by the marginal probability distribution $p(\mathbf{v})$ of visible variables
\begin{equation}
    \tilde{p}(\mathbf{v}) \approx p(\mathbf{v}) = \frac{Z_v(\mathbf{v})}{Z},
\end{equation}
where $Z_v(\mathbf{v})$ is the conditional partition function (see (\ref{eq:Z}))
\begin{eqnarray}
    Z_v(\mathbf{v}) &=& \mathrm{Tr}_h e^{-\hat{H}} = \nonumber \\
    &=& e^{\mathbf{bv}} \mathrm{Tr}_h \exp \sum_{j,j^{\prime} = 1}^m (c_{jj^{\prime}} + \sum_{i=1}^nv_iw_{ijj^{\prime}})\hat{h}^+_j\hat{h}_{j^{\prime}} = \nonumber \\
    &=& e^{\mathbf{bv}} \mathrm{det}(\mathbf{1} + e^{H_h}).
\end{eqnarray}
For this purpose we adjust the model parameters $\mathbf{b}, \mathbf{c}, \mathbf{w}$ so that the log-likelihood of the data set $\mathbf{v}_{data}$ would be maximum for a given values of model parameters (see (\ref{eq:lhood})). To maximize the log-likelihood we use the gradient ascent algorithm again:
\begin{subequations}
\label{eq:sqRBM_gradient_ascent}
\begin{eqnarray}
    \Delta \mathbf{b} = \eta\partial_{\mathbf{b}}\mathcal{L}(\mathbf{b},\mathbf{c},\mathbf{w}), 
\end{eqnarray}
\begin{eqnarray}
    \Delta \mathbf{c} = \eta\partial_{\mathbf{c}}\mathcal{L}(\mathbf{b},\mathbf{c},\mathbf{w}),
\end{eqnarray}
\begin{eqnarray}
\Delta \mathbf{w} = \eta\partial_{\mathbf{w}}\mathcal{L}(\mathbf{b},\mathbf{c},\mathbf{w}).
\end{eqnarray}
\end{subequations}

The ascent step for the visible layer bias $\mathbf{b}$ coincides with that of RBM (\ref{eq:rbm_grads}), as far as the visible subsystem is postulated to be purely classical. Consider now the ascent step for the weights $\mathbf{w}$:
\begin{eqnarray}
\label{eq:draft}
    \partial_{\mathbf{w}}\mathcal{L}(\mathbf{b}, \mathbf{c}, \mathbf{w}) &=& \frac{1}{N_{data}}\sum_{\mathbf{v}_{data}}\left(\partial_{\mathbf{w}}\log Z_v(\mathbf{v_{data}}) - \partial_{\mathbf{w}}\log Z\right)= \nonumber \\ &=&\frac{1}{N_{data}}\sum_{\mathbf{v}_{data}}\Bigg( Z_v^{-1}(\mathbf{v_{data}})\partial_{\mathbf{w}}Z_v(\mathbf{v_{data}}) \bigg. \nonumber \\
     &&\qquad\qquad\left.- Z^{-1}\sum_{\mathbf{v}}\partial_{\mathbf{w}}Z\right).
\end{eqnarray}
Let us evaluate separately the gradient of $Z_v(\mathbf{v_{data}})$:
\begin{eqnarray}
    \partial_{\mathbf{w}}Z_v &=& e^{\mathbf{b}\mathbf{v}}\partial_{\mathbf{w}}\mathrm{det}(\mathbf{1} + e^{H_h}) = \nonumber \\
    &=& e^{\mathbf{b}\mathbf{v}}\mathrm{Tr}_h\left(\mathrm{det}(\mathbf{1}+e^{H_h})\cdot(1 + e^{H_h})^{-1}\partial_{\mathbf{w}}(\mathbf{1}+e^{H_h}) \right) = \nonumber \\
    &=&Z_v\mathrm{Tr}_h\left( p(\hat{\mathbf{h}}^+\hat{\mathbf{h}}|\mathbf{v}) \mathbf{v}\hat{\mathbf{h}}^+\hat{\mathbf{h}}\right).
\end{eqnarray}
Substituting this in (\ref{eq:draft}) and assuming that there are $N_{data}$ samples of visible units in the input data set, we obtain
\begin{eqnarray}
    \partial_{\mathbf{w}}\mathcal{L}(\mathbf{b},\mathbf{c},\mathbf{w}) &=& \frac{1}{N_{data}}\sum_{\mathbf{v_{data}}}\mathrm{Tr}_h[p(\hat{\mathbf{h}}^+\hat{\mathbf{h}}|\mathbf{v}_{data})\mathbf{v}_{data}\hat{\mathbf{h}}^+\hat{\mathbf{h}}] \nonumber \\
    &&\qquad\,- \sum_{\mathbf{v}}\mathrm{Tr}_h[p(\hat{\mathbf{h}}^+\hat{\mathbf{h}}|\mathbf{v})\mathbf{v}\hat{\mathbf{h}}^+\hat{\mathbf{h}}].
\end{eqnarray}
For the gradient of the log-likelihood w.r.t. the hidden layer bias $\mathbf{c}$ one can proceed in the same way to obtain
\begin{eqnarray}
    \partial_{\mathbf{c}}\mathcal{L}(\mathbf{b},\mathbf{c},\mathbf{w}) &=& \frac{1}{N_{data}}\sum_{\mathbf{v}_{data}}\mathrm{Tr}[p(\hat{\mathbf{h}}^+\hat{\mathbf{h}}|\mathbf{v}_{data})\hat{\mathbf{h}}^+\hat{\mathbf{h}}] \nonumber \\
    &&\qquad\,- \sum_{\mathbf{v}}\mathrm{Tr}[p(\hat{\mathbf{h}}^+\hat{\mathbf{h}}|\mathbf{v})\hat{\mathbf{h}}^+\hat{\mathbf{h}}].
\end{eqnarray}
The updates of the biases $\Delta\mathbf{b},\Delta\mathbf{c}$ and weights $\Delta\mathbf{w}$ are proportional to the relative gradients with the proportionality factor $\eta$, which is the learning rate.

\subsection{Gibbs sampling}
During the training of both classical RBM and sqRBM one needs to evaluate the mean values of different quantities namely visible units, hidden units and their combinations. In practice it can be implemented by generating a large amount (representative set) of samples with some underlying probability distribution, which is not known explicitly, and then by taking the mean value over the generated samples. To do so one may apply the so-called Gibbs sampling \cite{Fischer2012}, which allows to produce samples from some joint probability distribution $(\mathbf{v},\mathbf{h})$. Basically it allows us to update the state of visible layer $\mathbf{v}$ for a given hidden layer state $\mathbf{h}$ and vice versa. This procedure is applicable and quite efficient because all hidden units are independent of each other, and so are the visible ones.

During the sqRBM training though we cannot apply Gibbs sampling straightforwardly. This is because of the $m \times m$ matrix introduction for the hidden layer instead of a $m$-dimensional vector with mutually independent components. In this case diagonalization can help with this obstacle. One has to include this step into the usual Gibbs sampling algorithm. Update of the visible layer for a given hidden state stays the same as for the classical RBM.

To sum up, in table \ref{tab:sqRBM} we suggest the pseudo-code for the numerical realization of sqRBM \cite{Fischer2012, Hinton2010}.
\begin{table}[b]
\caption{\label{tab:sqRBM}%
Pseudo-code of the sqRBM Training}
\begin{ruledtabular}
\begin{tabular}{ll}
$\mathrm{\textrm{1. Initialization of machine:}}$& \\
$\mathrm{\mathbf{w} \leftarrow \mathnormal{gauss(\mu = 0, \sigma = 0.01)}}$ &\\
$\mathrm{\mathbf{b}, \mathbf{c} \leftarrow \mathnormal{(0 ... 0)}}$& \\
$\textrm{2. Training procedure:}$& \\
$\mathrm{For\quad\mathbf{e}\quad in \quad\mathnormal{(\mathbf{1} ... \mathbf{E})} \quad do:}$&\\
$\qquad\mathrm{\mathbf{v^0} \leftarrow (\mathnormal{training \quad set})}$ &\\
$\qquad\mathrm{(\hat{\mathbf{h}}^+\hat{\mathbf{h}})^\mathbf{0} \leftarrow \mathnormal{\sigma_{m}}\left( \mathbf{v^0 \cdot w} + \mathbf{c} \right)}$ &\\
$\qquad \mathrm{For\quad \mathbf{k}\quad in \quad\mathnormal{(\mathbf{1} ... \mathbf{K})} \quad do:}$&\\
$\qquad\qquad\mathrm{Diagonalization\left[(\hat{\mathbf{h}}^+\hat{\mathbf{h}})^{\mathbf{k-1}}\right]}$&\\
$\qquad\qquad\mathrm{\mathbf{v^k} \leftarrow sample\left[\mathnormal{\sigma_{el}} \left( (\hat{\mathbf{h}}^+\hat{\mathbf{h}})^{\mathbf{k-1}} \cdot \mathbf{w} + \mathbf{b} \right)\right]}$&\\
$\qquad\qquad\mathrm{(\hat{\mathbf{h}}^+\hat{\mathbf{h}})^{\mathbf{k}} \leftarrow \mathnormal{\sigma_{m}} \left( \mathbf{v^k \cdot w} + \mathbf{c} \right)}$&\\
$\qquad\mathrm{\mathbf{w} \leftarrow \mathbf{w} + \eta\left(\langle \mathbf{v}\hat{\mathbf{h}}^+\hat{\mathbf{h}} \rangle^0 - \langle \mathbf{v}\hat{\mathbf{h}}^+\hat{\mathbf{h}} \rangle^k\right)}$&\\
$\qquad\mathrm{\mathbf{b} \leftarrow \mathbf{b} +  \eta\left(\langle \mathbf{v} \rangle^0 - \langle \mathbf{v} \rangle^k\right)}$&\\
$\qquad\mathrm{\mathbf{c} \leftarrow \mathbf{c} + \eta \left( \langle \hat{\mathbf{h}}^+\hat{\mathbf{h}} \rangle^0 - \langle \hat{\mathbf{h}}^+\hat{\mathbf{h}} \rangle^k\right)}$&
\end{tabular}
\end{ruledtabular}
\end{table}

Here $\mathbf{E}$ is the number of descent steps; $\mathbf{K}$ is the number of Gibbs sampling steps; $\mathnormal{\sigma_{el/m}}$ is the element-wise/matrix logistic (sigma) function; $\mathnormal{sample[...]}$ is the stochastic turning on of visible unit with the probability $\mathnormal{[...]}$; $\mathbf{\Delta ...}$ is the increment of $\mathbf{...}$ \cite{Hopfield:1982}.

\subsection{Persistent Contrastive Divergence}
  The exact evaluation of the gradient in (\ref{eq:sqRBM_gradient_ascent}) can be achieved by the Monte Carlo simulation. The positive part of the gradient involves the averaging of the single-particle density matrix (\ref{eq:FermiDirac}) and its moments over the training data samples $\mathbf{v}_{data}$. The negative part of the gradient  involves the averaging of the same quantities over  $\mathbf{v}$  with the model probability $p(\mathbf{v})$.  The latter requires to perform a separate Monte Carlo simulation via the Metropolis algorithm \cite{Hastings1970}  for each gradient ascent step which is impractical.  The way out is provided by the observation that when the learning rate  $\eta$  is small,  the model is changed only slightly, so that we can continue the Markov Chain from the state at the previous learning step. Provided that the change of  $\eta$  is sufficiently small, such a chain will follow close enough the probability distribution of the  model at the current set of parameters. In other words, we perform a single Metropolis simulation of the model. At each Monte Carlo step the negative part of the gradient is approximated by its value for the current configuration of the model. The positive part of the gradient is estimated by drawing a random training data sample $\mathbf{v}_{data}$ at each Monte Carlo step. Then the model parameters are updated according to (\ref{eq:sqRBM_gradient_ascent}).  This is called Persistent Chain (PC) \cite{Tieleman2008}. However the fluctuations of the gradient estimate may hinder the gradient ascent convergence. In order to reduce these fluctuations, one simulates not single but several independent concurrent Persistent Chains, and the model parameters increment is averaged over the chains. This is called Persistent Contrastive Divergence algorithm. 
  
\begin{table}[b]
\caption{\label{tab:persistent}%
Pseudo-code of Persistent Contrastive Divergence Training Procedure for sqRBM}
\begin{ruledtabular}
\begin{tabular}{ll}
$\mathrm{\textrm{1. Initialization of machine:}}$& \\
$\mathrm{\mathbf{w} \leftarrow \mathnormal{gauss(\mu = 0, \sigma = 0.01)}}$&\\
$\mathrm{\mathbf{b}, \mathbf{c} \leftarrow \mathnormal{(0 ... 0)}}$&\\
$\mathrm{For\quad\mathbf{i}\quad in \quad\mathnormal{(\mathbf{1} ... \mathbf{N_{pcd}})} \quad do:}$&\\
$\qquad\mathrm{\mathbf{v_{pcd}^0[i]} \leftarrow (\mathnormal{vector \, of \, random \, bits})}$&\\
$\qquad\mathrm{\mathbf{w_{pcd}^0[i]} \leftarrow p(\mathbf{v_{pcd}^0[i]})}$&\\
$\textrm{2. Training procedure:}$&\\
$\mathrm{For\quad\mathbf{e}\quad in \quad\mathnormal{(\mathbf{1} ... \mathbf{E})} \quad do:}$&\\
$\qquad\mathrm{\Delta \mathbf{w}  =  0}$&\\
$\qquad\mathrm{\Delta \mathbf{b} = 0}$&\\
$\qquad\mathrm{\Delta \mathbf{c} = 0}$&\\
$\qquad\mathrm{\mathbf{v_{data}} \leftarrow (\mathnormal{random \, instance \, of \, training \, set})}$&\\
$\qquad\mathrm{(\hat{\mathbf{h}}^+\hat{\mathbf{h}})_{data} \leftarrow \mathnormal{\sigma_{m}}\left( \mathbf{v_{data} \cdot w} + \mathbf{b} \right)}$&\\
$\qquad \mathrm{For\quad \mathbf{i}\quad in \quad\mathnormal{(\mathbf{1} ... \mathbf{N_{pcd}})} \quad do:}$&\\
$\qquad \qquad\mathrm{\mathbf{v_{trial}} \leftarrow (\mathnormal{generate \, trial \, configuration \, from \, \mathbf{v_{pcd}^{e-1}[i]})}}$&\\
$\qquad \qquad \mathrm{\mathbf{w_{trial}} \leftarrow p(\mathbf{v_{trial}})}$&\\
$\qquad\qquad\mathrm{If \, \mathbf{v_{trial}} \, is \, accepted \, based \, on \, Metropolis \, rule}$&\\
$\qquad \qquad \qquad \mathrm{for \, \mathbf{w_{trial}/w_{pcd}^{e-1}[i]} \, then\,\mathbf{v_{pcd}^{e}[i]} \leftarrow \mathnormal{\mathbf{v_{trial}}}}$&\\
$\qquad\qquad\mathrm{Else}$&\\
$\qquad \qquad \qquad \mathrm{\mathbf{v_{pcd}^{e}[i]} \leftarrow \mathnormal{\mathbf{v_{pcd}^{e-1}[i]}}}$&\\
$\qquad\qquad\mathrm{(\hat{\mathbf{h}}^+\hat{\mathbf{h}})_{\mathbf{pcd}}\mathbf{[i]} \leftarrow \mathnormal{\sigma_{m}} \left( \mathbf{v_{pcd}^{e}[i] \cdot w} + \mathbf{b} \right)}$&\\
$\qquad \qquad\mathrm{\Delta \mathbf{w} = \Delta \mathbf{w} + \mathbf{v_{data}}(\hat{\mathbf{h}}^+\hat{\mathbf{h}})_{\mathbf{data}} -  \mathbf{v_{pcd}^{e}[i]} (\hat{\mathbf{h}}^+\hat{\mathbf{h}})_{\mathbf{pcd}}\mathbf{[i]}}$&\\
$\qquad \qquad\mathrm{\Delta \mathbf{b} = \Delta \mathbf{b} +  \mathbf{v_{data}} - \mathbf{v_{pcd}^{e}[i]}}$&\\
$\qquad \qquad\mathrm{\Delta \mathbf{c} = \Delta \mathbf{c} +(\hat{\mathbf{h}}^+\hat{\mathbf{h}})_{\mathbf{data}}- (\hat{\mathbf{h}}^+\hat{\mathbf{h}})_{\mathbf{pcd}}\mathbf{[i]}}$&\\
$\qquad \mathrm{\mathbf{w} \leftarrow \mathbf{w} + \eta\Delta \mathbf{w}}$&\\
$\qquad \mathrm{\mathbf{b} \leftarrow \mathbf{b} + \eta\Delta \mathbf{b}}$&\\
$\qquad \mathrm{\mathbf{c} \leftarrow \mathbf{c} + \eta\Delta \mathbf{c}}$&
\end{tabular}
\end{ruledtabular}
\end{table}
  
\section{Tests}
\begin{figure}
\begin{minipage}[h]{\linewidth}
    \center{\includegraphics[width=0.65\linewidth]{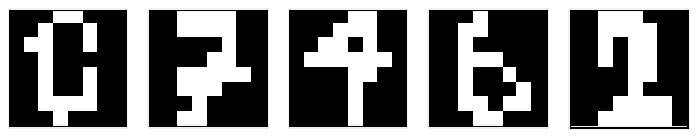} \\ (a)}
\end{minipage}
\vfill
\begin{minipage}[h]{\linewidth}
    \center{\includegraphics[width=0.9\linewidth]{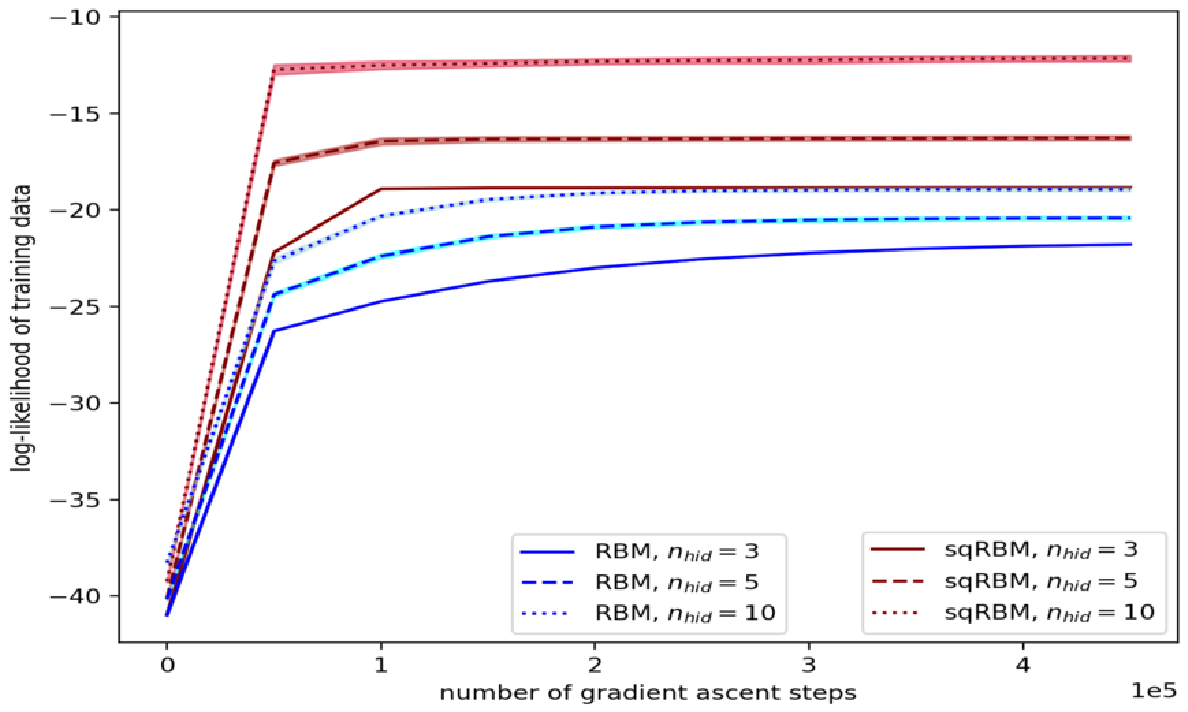} \\ (b)}
\end{minipage}
\vfill
\begin{minipage}[h]{\linewidth}
    \center{\includegraphics[width=0.9\linewidth]{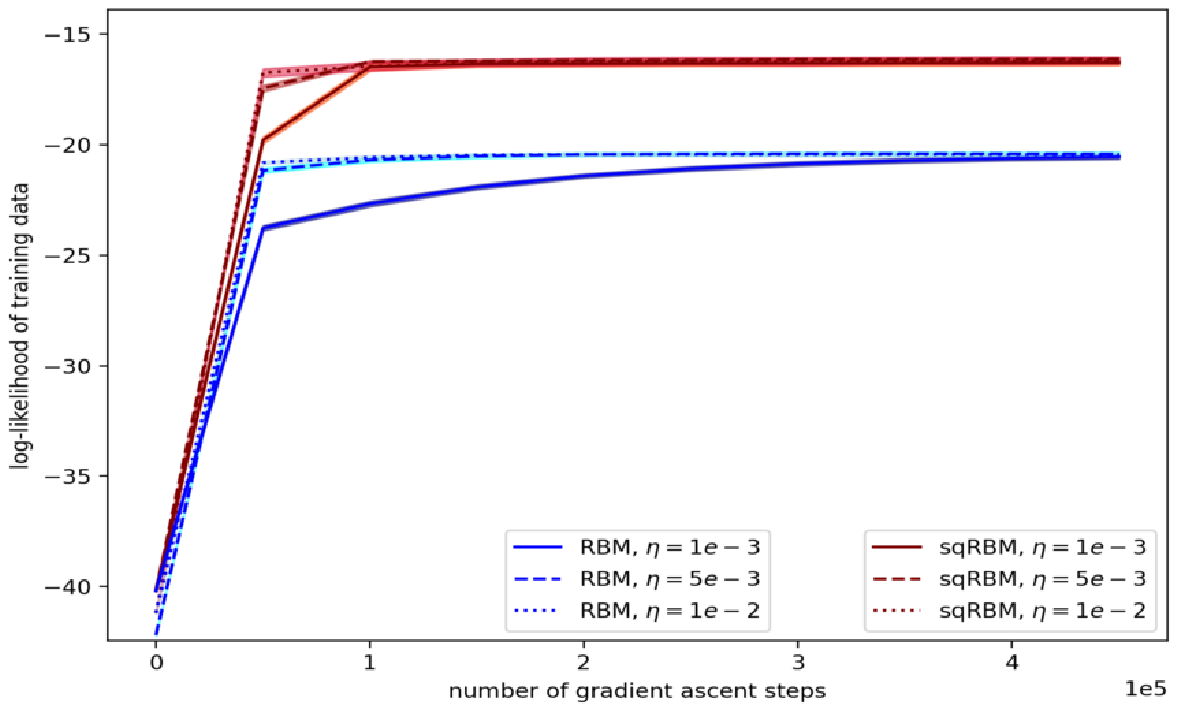} \\ (c)}
\end{minipage}
    \caption{Log-likelihood vs. the number of ascent steps for the training on the OptDigits set (a). Training curves for different $n_{hid}$ (b) and $\eta$ parameters are shown.}
    \label{fig:opt_set}
\end{figure}

We perform the experiments with RBM and sqRBM by training the machines on two familiar data sets. The first one is a set of $4 \times 4$ square pictures with random black-and-white \textit{Bars\&Stripes} (see figure \ref{fig:bs_set}a). The second one is \textit{Optdigits} \cite{OptDigits} (see figure \ref{fig:opt_set}a). This is a set of 5620 $8 \times 8$ originally grayscale samples of handwritten digits, which we simplified to black-and-white ones (see figure \ref{fig:opt_set}). In each test the number of visible neurons was fixed to the number of pixels in one image.
\subsection{Training by Contrastive Divergence}
The first experiment was conducted in order to compare the performances of RBM and sqRBM with use of simple Contrastive Divergence algorithm with one Gibbs sampling step ($CD_1$). The measure of training quality was the log-likelihood of the training data set. For estimating the log-likelihood we used annealed importance sampling (AIS) \cite{Neal2001}. We meant to conduct the numerical experiments not only to conclude whether the sqRBM can learn better then the RBM, but also to study the hyperparameters dependence such as number of hidden units and learning rate.

\begin{figure}
\begin{minipage}[h]{\linewidth}
    \center{\includegraphics[width=0.65\linewidth]{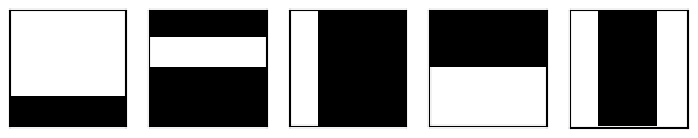} \\ (a)}
\end{minipage}
\vfill
\begin{minipage}[h]{\linewidth}
    \center{\includegraphics[width=0.98\linewidth]{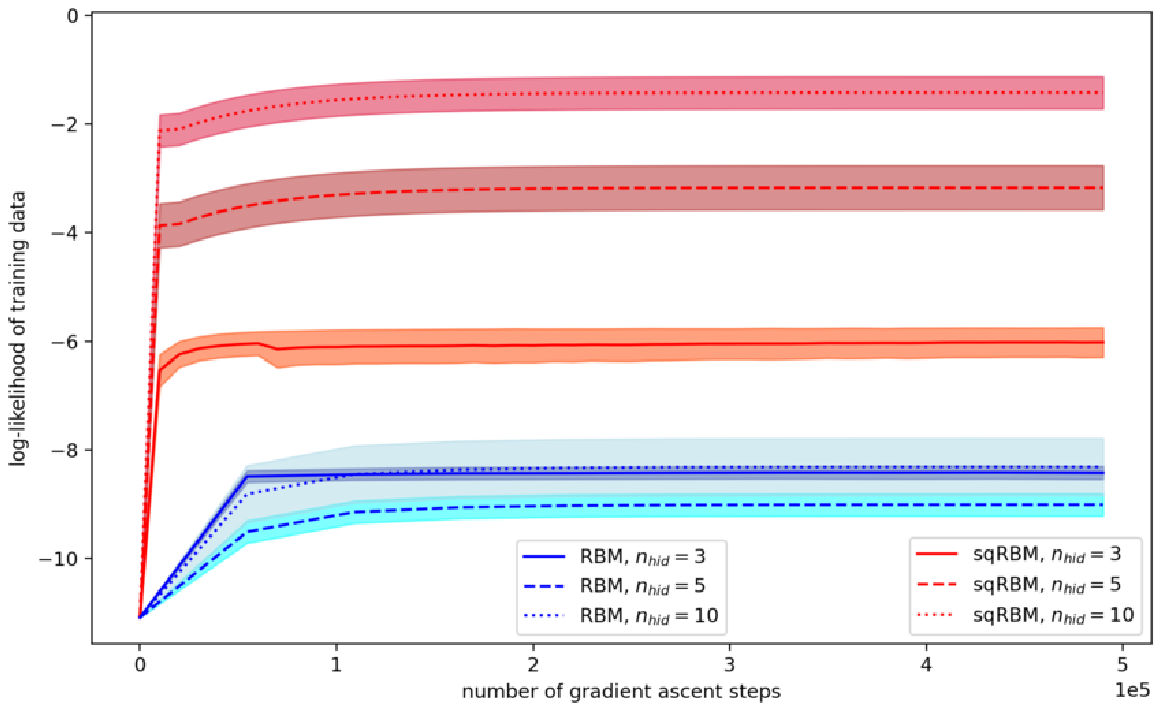} \\ (b)}
\end{minipage}
\vfill
\begin{minipage}[h]{\linewidth}
    \center{\includegraphics[width=0.98\linewidth]{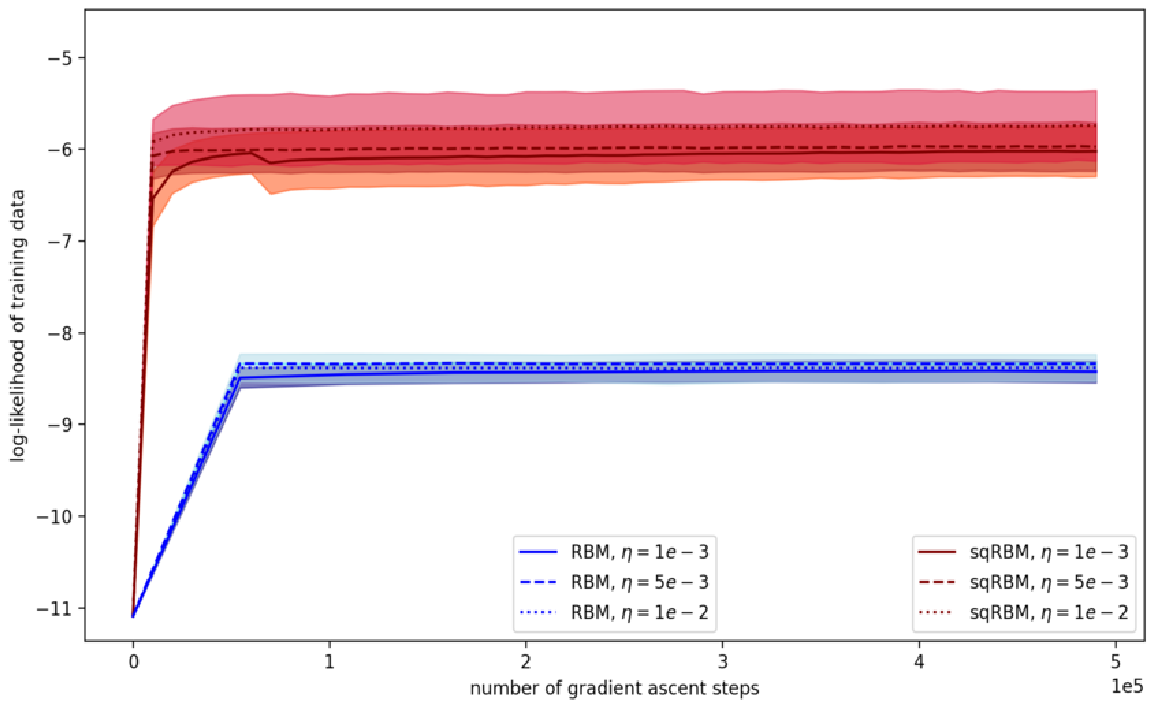} \\ (c)}
\end{minipage}
    \caption{Log-likelihood vs. the number of ascent steps for the training on the set of bars and stripes (a). Training curves for different $n_{hid}$ (b) and $\eta$ parameters are shown.}
    \label{fig:bs_set}
\end{figure}

The results of the training on the Bars\&Stripes and the OptDigits sets are shown at the figures \ref{fig:bs_set} and \ref{fig:opt_set} respectively. For both sets the sqRBM demonstrates not only higher log-likelihood achieved, but also more rapid saturation. It is important to note that such a tendency doesn't depend on the hyper-parameters. All the curves are provided with the error area, which presents the dispersion due to the random initial weights of the machines.

\subsection{Training by Persistent Contrastive Divergence}

\begin{figure}
    \centering
    \includegraphics[scale=0.43]{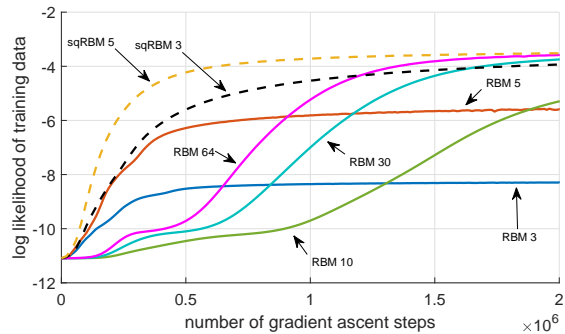}
    \caption{Modelling the 4x4 Bars\&Stripes dataset with RBM and sqRBM at different numbers of hidden units. The type of model and the number of hidden units are indicated by arrows. }
    \label{fig:bnb_logl}
\end{figure}

In figure \ref{fig:bnb_logl} the results of training RBM and sqRBM on the 4x4 Bars\&Stripes dataset are presented. It shows that making the hidden units quantum leads to significant improvement of the learning capabilities of the model. sqRBM was trained with the constant learning rate $\eta=0.001$. In the case of RBM, $\eta=0.001$ for the number of hidden units $\leq5$, and $\eta=0.0001$ for higher number of hidden units. In all cases the minibatch size (number of concurrent PC) is 100. The training dataset was 40000 random instances of the Bars\&Stripes.

\begin{figure}
    \centering
    \includegraphics[scale=0.39]{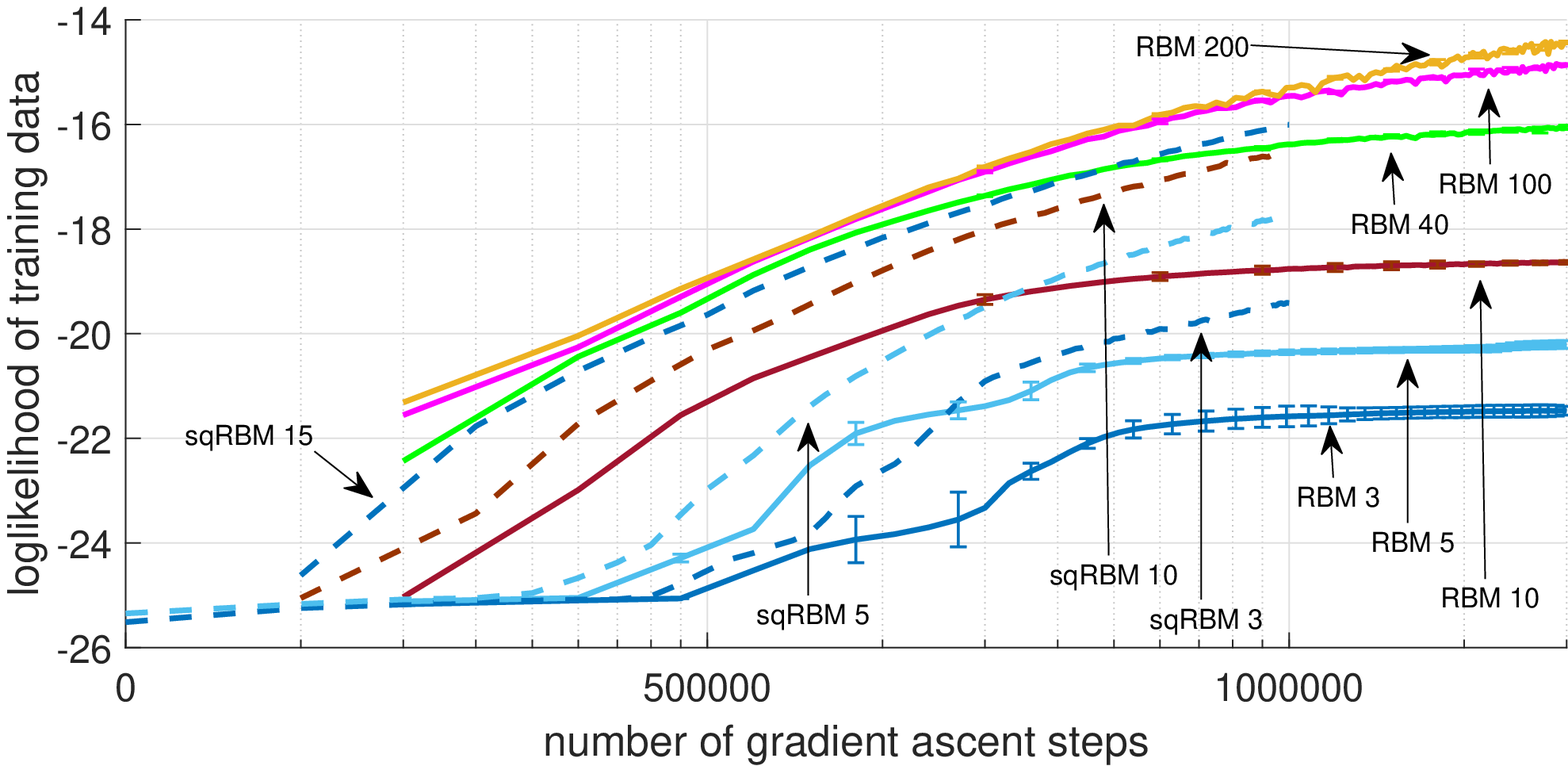}
    \caption{Modelling the Optdigits dataset with RBM and sqRBM at different numbers of hidden units. The type of model and the number of hidden units are indicated by arrows. }
    \label{fig:opt_logl}
\end{figure}

In figure \ref{fig:opt_logl} the results of training of RBM and sqRBM on the Optdigits dataset are presented. In all the cases $\eta=0.001$ was employed. The minibatch size is 100. RBM curves are additionally supplemented with the error bars, whereas the error bars for the sqRBM curves appeared to be comparatively indistinguishable. Here we also observe the improvement due to the quantumness of the hidden units. The convergence rate of sqRBM is also faster than of RBM. 

We conclude with the observation that the sqRBM with $N$ quantum hidden units tends to learn better that the RBM with $N^2$ classical hidden units  

\begin{figure}
    \centering
    \includegraphics[width=\linewidth]{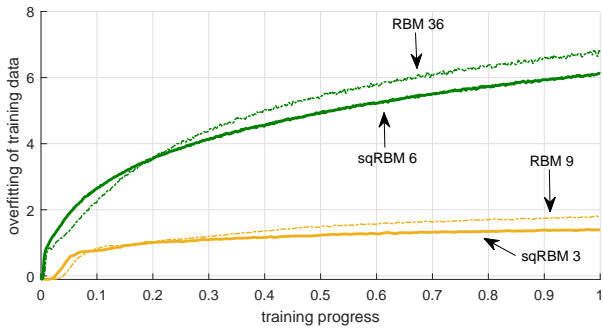}
    \caption{Overfitting of the RBM and sqRBM models on the Optidits dataset. The "training progress" are the gradient ascent steps up to the saturation of log-likelihood on the training data. The step numbers are normalized to be 1 at the point of saturation of the training log-likelihood. }
    \label{fig:opt_overfitting}
\end{figure}

\subsection{Overfitting}
The overfitting (i.e. degradation of the model predictive capability) may be estimated as the likelihood drop when the model operates the data it has not seen before. Here we present the results of such an estimation. The Optdigits data set was partitioned into two parts.The first one is comprised of the first 2800 samples. And the second one, comprised of 5600 samples, was considered as the validating subset. The machines learned on the training subset, and the corresponding log-likelihood of the training subset was calculated. Then the log-likelihood of the validating subset was calculated. The amount by which the log likelihood of the validating subset decreases is the measure of the overfitting: the higher the worse.  In figure \ref{fig:opt_overfitting}  we present the overfittings of sqRBM and of RBM models with a quadratically larger number of hidden  units (see the end of the previous section), as a function of the training progress. The latter is defined as follows. As the models are trained, the log-likelihood first starts to grow then saturates at a certain level (i.e. the model no longer learns from the training data set). As the saturation is achieved, the training is stopped. This corresponds to a certain (maximal) number of gradient ascent steps. For each model, we "normalize" the training progress by dividing the current number of the gradient ascent step by the maximal number before saturation. We see that sqRBM shows slightly better results than RBM.

\section{Conclusion}

We proposed and examined a semi-quantum version of the Restricted Boltzmann Machine with classical visible units and fermionic hidden units which we called semi-quantum RBM (sqRBM). The presented sqRBM model inherits simple and effective trainability of the classical RBM. In particular this model can be trained using the Gibbs sampling with including diagonalization w.r.t. the hidden units as an additional step. 

 At the same time introduction of quantum hidden degrees of freedom makes the model sufficiently more flexible than the purely classical one. Moreover sqRBM avoid overfitting better than the classical RBM. It was confirmed during the numerical experiments with the use of two standard data sets, namely Bars\&Stripes and OptDigits. As a performance measure we used the log-likelihood estimation via annealed importance sampling. Future work should investigate whether this success stays for larger data sets.
 
 The distinctive features of our model (hybrid classical/quantum system and fermionic quantum part) makes it an interesting option for the development of quantum-inspired machine learning algorithms.
 
\section*{Acknowledgements}
The work of Ya.S.L. is supported by the Russian Foundation for Basic Research, through the joint RFBR and CNR project No. 20-52-7816.

\bibliography{main}

\end{document}